\documentclass[sigconf]{acmart}
\AtBeginDocument{%
  }

\usepackage{tikz}
\usepackage{balance}
\usepackage{pgfplots}
\pgfplotsset{compat=1.18}

\copyrightyear{2026}
\acmYear{2026}
\setcopyright{usgov}
\acmConference[SIGCSE Virtual 2026]{Proceedings of the 2nd ACM Virtual Global Computing Education Conference V.1}{November 12--15, 2026}{Virtual Event, USA}
\acmBooktitle{Proceedings of the 2nd ACM Virtual Global Computing Education Conference V.1 (SIGCSE Virtual 2026), November 12--15, 2026, Virtual Event, USA}
\acmDOI{10.1145/3795867.3830988}
\acmISBN{979-8-4007-2506-7/2026/11}

\begin{document}

\title{Ethics Integration in CS Theory Courses}

\author{Ryan E. Dougherty}
\email{ryan.dougherty@westpoint.edu}
\orcid{0000-0003-1739-1127}
\affiliation{%
  \institution{United States Military Academy}
  \city{West Point}
  \state{New York}
  \country{USA}
}


\renewcommand{\shortauthors}{Dougherty}

\begin{abstract}
The integration of ethics into computing curricula has gained substantial momentum, based on foundational work by the Embedded EthiCS movement, ACM/IEEE curricular recommendations, and a growing literature on ethics modules in courses on artificial intelligence, software engineering, human-computer interaction, and data science. One foundational area of the undergraduate computing curriculum remains absent from these efforts: CS theory. Courses on automata, computability, and complexity are widely perceived as too abstract or value-neutral for sustained ethical engagement, and the published record on ethics integration largely reflects this perception. In this position paper, we argue that theory courses are not merely compatible with ethics integration but offer a distinctive and underutilized perspective that complements theory itself. We synthesize the embedded-ethics literature, examine the standard objections to ethics in theory, and identify five avenues through which canonical theory results connect directly to substantive ethical questions: undecidability and the limits of automated reasoning, complexity-theoretic privacy and fairness, formal languages and pattern matching, the rhetoric of problem reductions, and randomization as an ethical primitive. We propose lightweight integration strategies grounded in existing pedagogical literature and respond to the most common instructor objections.
\end{abstract}



\maketitle

\section{Introduction}

Over the past decade, the computing education community has invested
substantial effort in integrating ethical reasoning into the
undergraduate curriculum. 
The \emph{Embedded EthiCS} program at Harvard~\cite{grosz2019embedded} popularized a model in which ethics specialists and CS instructors co-design short, contextually relevant interventions embedded within existing technical courses rather than relegating ethics to a single separate offering.
Related work has explored distributed or embedded approaches.
The ACM/IEEE/AAAI \emph{Computer Science Curricula 2023} guidelines~\cite{cs2023} now treat Society, Ethics, and the Profession as a first-class knowledge area, while the ACM Code of Ethics and Professional Conduct~\cite{acmcode2018} articulates corresponding professional responsibilities.
Syllabus studies such as \citet{fiesler2020we} identify more than one hundred standalone technology-ethics courses, and a growing literature reports ethics integration in artificial intelligence~\cite{burton2018teach,saltz2019integrating}, data science~\cite{saltz2019integrating}, and human-computer interaction~\cite{skirpan2018ethics}.

Theory of Computing (ToC) courses, defined here as those containing a substantial portion of automata theory, formal languages, computability, and/or complexity~\cite{dougherty_survey_2024}, has received comparatively little attention.
These courses are required in many undergraduate CS programs, yet instructors have few tailored integration resources.
Surveys of integrated modules~\cite{fiesler2020we,saltz2019integrating} identify essentially no interventions centered on canonical ToC results, and widely used theory textbooks~\cite{sipser2013introduction,hopcroft2007introduction} contain no sustained ethics content.

We argue that the perception of theory as too abstract~\cite{braught2025too} or too distant from practice for ethical engagement is mistaken.
ToC offers a unique perspective: computability and complexity results constrain what algorithmic systems can promise, while reductions and formal models train students to ask what a transformation preserves and omits.
When these results are taught without inviting students to examine their implications, graduates may possess the vocabulary of computational limits without having practiced applying it to claims about verification, privacy, security, automated decision-making, or algorithmic inevitability/necessity.
Treating those connections as outside the scope of the course is therefore a curricular framing choice rather than the absence of framing.

Our position is that ToC courses are simultaneously applicable and important areas for ethics integration in the undergraduate CS curriculum. 
Our intended audience consists of ToC instructors and CS curriculum designers. 
We make four contributions:
\begin{itemize}
    \item We synthesize the embedded-ethics literature and document the relative absence of canonical ToC content.
    \item We examine three recurring objections to integration and connect them to empirically documented barriers faced by computing educators.
    \item We identify five avenues through which standard ToC results support substantive ethical reasoning without requiring new technical units.
    \item We propose lightweight integration strategies, respond to implementation concerns, and close with a research and curricular agenda.
\end{itemize}

\section{Existing Computing Ethics Integration}
\label{sec:landscape}

Modern movements to integrate ethics into computing courses can be traced to a confluence of external pressures in the mid-2010s: high-profile failures of deployed algorithmic systems~\cite{oneil2016weapons,Lebovits2019automating}, calls from within CS to revisit professional responsibility~\cite{acmcode2018}, and a growing recognition that standalone ethics courses did not reliably transfer to students' technical practice~\cite{grosz2019embedded}. 
The \emph{Embedded EthiCS} model by \citet{grosz2019embedded} responded to these concerns with a specific proposal: rather than teaching ethics in isolation, philosophers and computing professionals would design short, contextually relevant modules within existing technical courses.
The purpose was that ethical reasoning would be encountered alongside relevant technical material. 
There have been many adaptations and an active community of practice~\cite{fiesler2020we}.
Two parallel developments reinforced the integration, namely the ACM/IEEE/AAAI \emph{Computer Science Curricula 2023} curricular guidelines \cite{cs2023}, and the revised ACM Code of Ethics and Professional Conduct~\cite{acmcode2018}. 

The transformation reflects a clear disciplinary skew. 
For example, Fiesler, Garrett, and Beard's analysis of more than 100 technology ethics syllabi~\cite{fiesler2020we} found that the topics most frequently addressed cluster around law and policy, privacy and surveillance, and the social implications of artificial
intelligence and data science.
\citet{saltz2019integrating} documented a similar concentration in ML courses, finding that only 12\% of technical ML courses contained any ethics-related content.
\citet{skirpan2018ethics} reported on integration in HCI; \citet{burton2018teach} situated their science-fiction-based interventions in AI courses. 
Software engineering and introductory programming have also received sustained attention, the latter through exercises adapted to CS1/2~\cite{cohney2021ethics}. 
A recent systematic review of ACM computing-education ethics publications~\cite{brown2024teaching} documented 100 papers across many computing education venues through 2022 and found the same disciplinary concentration in ML, data science, HCI, software engineering, and intro programming.

Recent curriculum-level studies also show that ethics requirements remain uneven, although their measures differ. 
\citet{weichert2025ethics} found that 33\% of the 250 programs they reviewed required a standalone computing-ethics course; their analysis explicitly excluded ethics content embedded in technical courses. 
\citet{barkhuff2026mapping} used a sample of 100 U.S. bachelor's programs, and counting both standalone and embedded instruction found that 55\% required ethics instruction in some capacity. 
These estimates show that both the prevalence and the location of required ethics instruction are uneven.

\begin{figure*}[t]
\centering
\begin{tikzpicture}
\begin{axis}[
    xbar,
    width=1.5*\columnwidth,
    height=9cm,
    bar width=11pt,
    enlarge y limits=0.08,
    xmin=0,
    xmax=30,
    xlabel={Number of modules},
    xlabel style={font=\small},
    symbolic y coords={toc,crypto,dm,sec,pl,ds,introcs,algs,hci,aiml,sys},
    ytick=data,
    yticklabels={%
                {Systems/Architecture/Networks},
                {AI/ML/NLP},
                {HCI/visualization},
                Algorithms,
        Introductory CS,
        Data Science,
        Programming languages,
        Security,
        Discrete mathematics,
        Cryptography,
        Theory of computing},
    yticklabel style={font=\small},
    nodes near coords,
    nodes near coords style={font=\footnotesize},
    nodes near coords align={horizontal},
    tick style={draw=none}]
\addplot[fill=black!75, draw=none] coordinates {
    (27,sys)
    (24,aiml)
    (12,hci)
    (11,algs)
    (10,introcs)
    (10,ds)
    (8,pl)
    (4,sec)
    (3,dm)
    (2,crypto)
    (0,toc)};
\end{axis}
\end{tikzpicture}
\caption{Distribution of Embedded EthiCS modules at Harvard
by course's main CS subfield, Spring 2017 through Fall
2023~\cite{embeddedethicsharvard,grosz2019embedded}. Counts reflect
modules listed in the program's public repository, classified by the
home subfield of the course in which the module was embedded.
Modules in courses that span multiple subfields (e.g., cryptography) were assigned
to their primary identification in the Harvard dataset.
}
\label{fig:embedded-ethics-distribution}
\end{figure*}
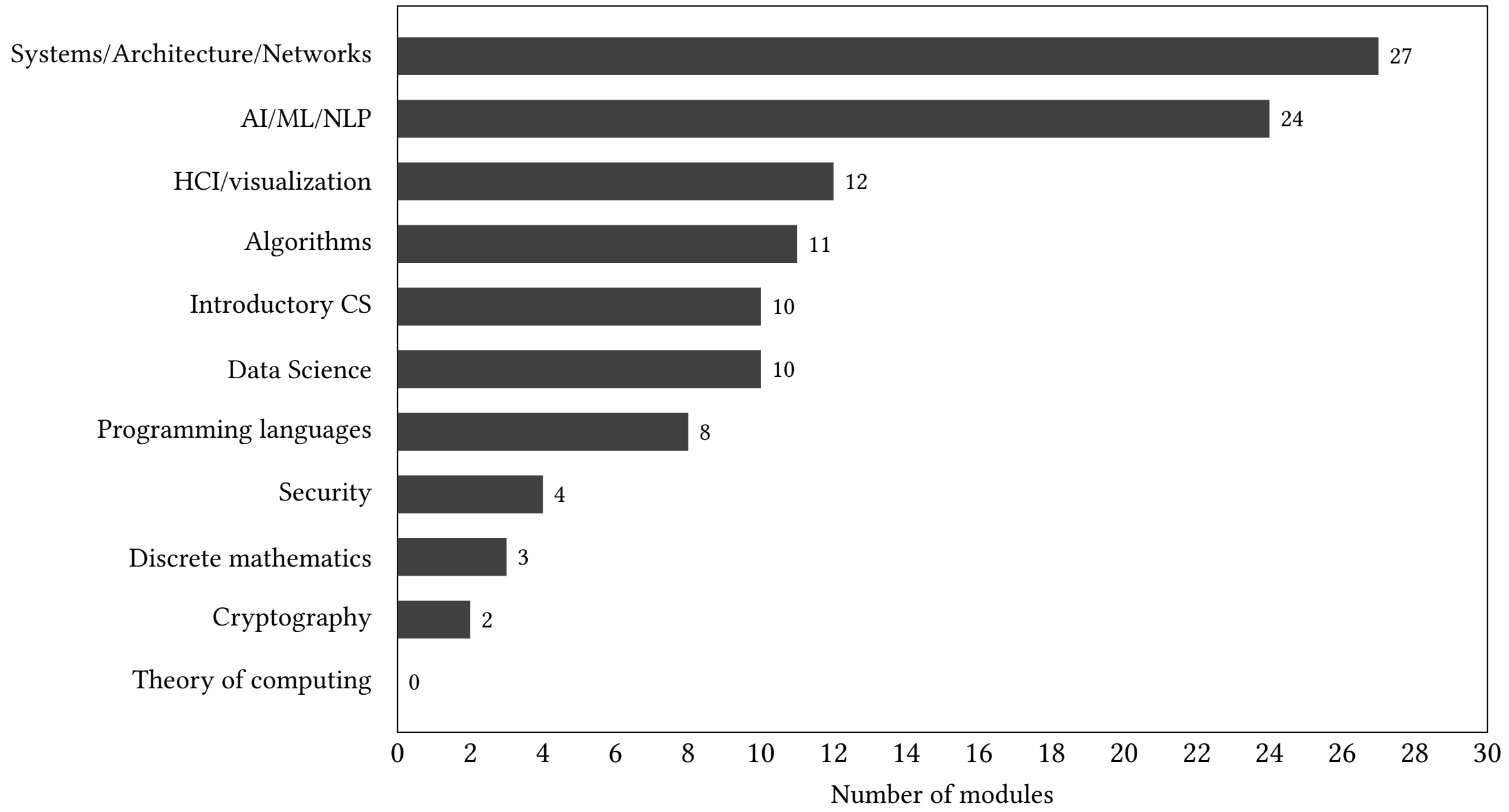

This skew is visible in the Embedded EthiCS program: Figure~\ref{fig:embedded-ethics-distribution} shows the distribution of modules by the CS subfield of the course in which each module was embedded, across seven academic years~\cite{embeddedethicsharvard}. 
Every major undergraduate CS subfield is represented with the single exception of ToC, which is not represented at all. 
The closest cases are modules in cryptography and in courses on algorithms or matching mechanisms, none of which engage the material that constitutes the standard undergraduate ToC course~\cite{dougherty_survey_2024}. 
We are not aware of a comparable empirical gap for any other major undergraduate CS subfield.

The ToC course is not represented in this literature in any comparable way. 
The Fiesler et al. dataset contains no theory-based syllabi.
Across the bounded sources examined here, we found no published intervention whose principal technical object was a canonical result such as the pumping lemma, Rice's theorem, or an NP-completeness reduction. 
This is a dataset-limited observation rather than the result of an exhaustive systematic search.
Standard undergraduate theory textbooks~\cite{sipser2013introduction,hopcroft2007introduction,arora2009computational} contain no ethics content, no exercises that frame technical results in social terms, and no instructor-facing material suggesting that such framing would be appropriate. 
The CS2023 report~\cite{cs2023} lists Society, Ethics, and the Profession as a knowledge area to be distributed across the curriculum but offers no specific guidance for theory courses, and the Society and Professional Practice topics in earlier curricular volumes~\cite{cs2013} were similarly silent.

\section{Why Has Theory Been Left Out?}
\label{sec:objections-fail}

Three arguments recur in informal discussions of why ethics integration has bypassed ToC: that the subject is too \emph{abstract} to support ethical engagement, too \emph{value-neutral} for ethical content to attach to its results, and too \emph{time-constrained} to accommodate non-technical material. 
Each has some plausibility, but on further examination rests on assumptions that the embedded-ethics literature has already shown to be unreliable in comparably situated courses.

\subsection{The Abstraction Argument}
\label{sec:obj-abstract}

The abstraction argument holds that ToC operates at a level of generality at which ethical questions cannot be sensibly posed: a Turing machine is not a hiring system, a regular language is not a content filter, and the gap between the formal object and any deployed artifact is too wide to bridge in a classroom hour. 
Here, ethics belongs to courses whose technical content is closer to practice.

The argument misidentifies what abstraction does. 
Abstractions are not ethically inert but rather are choices about which features of a situation are to be preserved and which to discard, and those choices encode normative commitments~\cite{selbst2019fairness,green2021algorithmic}. 
\citet{selbst2019fairness} constructed five ``abstraction traps'' in fair-ML practice, each of which arises when a technical formalism is treated as ethically neutral when in fact it has ruled out considerations a fair system would need. 
Their analysis applies a stronger argument to the abstractions of theory: the decision to model a problem as language recognition, to identify two problems via many-one reduction, or to treat a randomized algorithm as ``essentially'' deterministic are choices that propagate into every downstream system. 
Far from being a reason to exclude theory from ethics integration, the centrality of abstraction is the reason to include it. 
Theory is the course in which students learn to make these choices; it is therefore the natural point at which to learn to interrogate them.

\subsection{The Value-Neutrality Argument}
\label{sec:obj-neutral}

The value-neutrality argument is a stronger form of the same claim: not merely that theory's abstractions are ethically distant, but that its results are ethically empty. 
For example, the halting problem is a theorem about Turing machines (an abstract computing model); P versus NP is a question about resource bounds. 
These results are ``mathematics,'' which is inherently value-free.

This argument is in tension with the practice of ToC itself. 
Initial work in differential privacy~\cite{dwork2014algorithmic} treats privacy as a technical necessity motivated by ethical concern.
The impossibility results within algorithmic fairness~\cite{kleinberg2017inherent,chouldechova2017fair} are presented as findings about what fair systems can and cannot achieve. 
These are not applications of theory to ethics layered on after the fact, but rather are theoretical results whose content is essentially ethical. 
To present complexity theory to undergraduates as though it had no contact with these developments is not neutrality, but a curation choice.
By selecting a particular subset of the field's results, this choice conveys a particular and possibly contestable picture of what theory's essence is. 
The choice to teach theory as ethically empty is itself an ethical choice.

\subsection{The Time-Constraint Argument}
\label{sec:obj-time}

The time-constraint argument is the most concrete and very common: theory courses are technically dense, the syllabus is already ``overcrowded'', and any minute spent on ethics is a minute lost from material students need to pass the course. 
This argument deserves serious engagement because the underlying observation is correct. 
Theory courses are dense, and many already cut topics that earlier generations of instructors taught.

The argument fails not because the constraint is illusory but because it applies symmetrically across the curriculum. 
Machine learning, software engineering, human-computer interaction, and CS1/2 courses all face comparable time pressure, and all have nonetheless accommodated ethics integration~\cite{grosz2019embedded,saltz2019integrating,skirpan2018ethics}.
The integration literature is unified in finding that effective modules are short (usually a single class session, sometimes less) and that they exploit the technical material around them instead of displacing them. 
A fifteen-minute discussion of the implications of Rice's theorem after the theorem has been proved does not displace the proof itself. 
The time-constraint argument, applied consistently, would bring unified ethics integration across CS. 
The fact that it is offered specifically about theory suggests it is operating as a proxy for the abstraction or value-neutrality arguments above.

\section{Five Avenues for Theory Ethics Integration}
\label{sec:avenues}

\emph{How} should integration in theory courses proceed?
We identify five avenues through the ToC curriculum that connect to substantive ethical questions and to a theorem already present in standard syllabi. 
We do not claim that this list is exhaustive (and in fact is very much not), but that it is sufficient to falsify the position that theory has nothing distinctive to contribute to ethics integration.

\subsection{Computability/Automated Reasoning Limits}
\label{sec:av-computability}

The first follows directly from the central computability results: the undecidability of the halting problem and its generalization in Rice's theorem~\cite{rice1953classes,sipser2013introduction}. 
The latter establishes that no algorithm can decide any non-trivial semantic property of arbitrary programs, a result that bounds, in principle, what any program-analysis or program-verification system can promise. 
The ethical application of this result is direct; automated content moderation~\cite{gillespie2020content}, automated code review, and the verification of ``aligned'' behavior in AI systems~\cite{russell2019human} regularly treat these as engineering problems whose solutions are a matter of sufficient effort. 
Rice's theorem says their guarantees can never be total, and that any deployment that treats them as total is making a false claim.
Any discussion about building such systems must then inevitably intersect related ethical questions in other fields: economics, politics, sociology, etc. 

The pedagogical use-case is straightforward. 
After proving Rice's theorem, an instructor can spend a single class segment posing questions such as: which contemporary systems make claims that, if taken at face value, would require deciding a non-trivial semantic property of arbitrary programs? 
Students who have internalized the theorem and application are positioned to evaluate such claims, and the exercise teaches a transferable habit of reading marketing language against theoretical limits. 

\subsection{Complexity as an Ethical Primitive}
\label{sec:av-complexity}

The second concerns computational hardness itself (e.g., NP completeness), typically introduced in the latter half of a standard ToC course. 
Hardness is normally presented with some resignation, but in a substantial body of contemporary work it operates as a resource and/or an opportunity.
Modern cryptography is built on the assumed hardness of problems like integer factorization and discrete logarithm~\cite{katz2014introduction}.
Its ability to protect communication, financial transactions, and personal data depends on the existence of hardness. 
The ethical content here is clear: a world in which P = NP is a world without public-key cryptography, and therefore a world with a qualitatively different relationship between individuals and institutions that hold their data.

This reframing connects the standard P-versus-NP discussion to questions students already care about. 
It also creates a more important ethical perspective.
For example, approximation algorithms ask what ``good enough'' means when the objects being approximated are people (e.g., in scheduling, allocation, or matching contexts).
Parameterized complexity asks which features of a problem make it tractable, and therefore which features a system designer can manipulate to alter what is computationally feasible. 
The pedagogical use-case here is to follow the proof of an NP-completeness result with a brief framing question: if this problem is hard, who benefits from its hardness, and who is harmed by it? 
The question admits productive answers across scheduling, routing, allocation, and verification problems that already appear in standard exercises.

\subsection{Formal Languages and the Politics of Pattern Matching}
\label{sec:av-languages}

The third is based on (usually) the earliest material of the standard theory course: regular languages and finite automata.
This material is often the technically least demanding section of the course, and is sometimes treated by instructors as preliminary. 
However, its ethical implications are inversely proportional to its difficulty. 
Every content filter, rule-based email spam classifier, input validation function, and signature-based intrusion-detection system is a decision procedure for testing membership in some formal language. 
The expressiveness of the language family directly determines what such a system can and cannot detect.

The ethical questions follow almost immediately. 
A regular-language content filter cannot, in principle, recognize patterns requiring unbounded counting or matched delimiters; a context-free filter can recognize more complex patterns but not all. 
The choice of language family is therefore a choice about which kinds of speech the filter can act on, which means it is also a choice about which kinds of speech will be over-blocked (false positives) and which will be under-blocked (false negatives). 
These distributional consequences can fall unevenly across user populations~\cite{sap2019risk,davidson2019racial}, and the choice of formalism is the center from where the distribution emerges.
The pedagogical use-case is after presenting the pumping lemma, ask students to construct a regular language that a real content-moderation system would need to recognize, and to identify what the lemma rules out. 
The exercise reinforces the technical content while making its social consequences accessible.

\subsection{The Rhetoric of Reductions}
\label{sec:av-reductions}

The fourth is the one we believe is most distinctive to theory and least accessible from any other course. 
\emph{Reductions} show that one problem can be transformed into another, and is a very useful tool for theoreticians to establish the hardness, decidability, or equivalence of problems. 
Reduction is also, although not under that name, the principal rhetorical move by which contemporary computing systems are justified.
For example, deploying an algorithmic hiring system is asserting that hiring reduces to classification; deploying an automated recidivism predictor is to assert that the question of whether a person will reoffend reduces to a prediction problem on a feature vector~\cite{barocas2023fairness}; and deploying a content-moderation classifier is to assert that the distinction between protected and prohibited speech reduces to language recognition.

These reductions are not technically ``wrong'' as a reduction is a mathematical object. 
But the question of whether the reduction \emph{preserves what matters} is a question the formalism does not by itself answer. 
A reduction from hiring to classification preserves whatever the classification's loss function captures and discards whatever it does not; the choice of loss function is the locus at which substantive normative content enters the system, and it enters under the technical-sounding guise of a reduction~\cite{passi2019problem}. 
We propose that undergraduate ToC courses begin treating reductions as
\emph{rhetorical} as well as mathematical objects, that students learn to ask of every reduction they construct or encounter what the reduction preserves and what it loses. 
This is the avenue at which the abstraction objection from Section~\ref{sec:obj-abstract} is most concretely answerable: the abstraction trap that \citet{selbst2019fairness} identify in fair-ML practice is, here, a question about what a reduction preserves.

The pedagogical weight is to assign, alongside a standard reduction exercise, a short written commentary in which the student articulates a hypothetical real-world deployment that depends on the reduction and identifies one feature the reduction does not preserve. 
This requires no additional class time and reuses an existing exercise.

\subsection{Randomization as an Ethical Primitive}
\label{sec:av-randomization}

The fifth concerns randomization, typically introduced in theory courses containing randomized algorithms, the class \textbf{BPP}, or in some curricula through probabilistic finite automata. 
Randomization is normally presented as a tool for efficiency, trading deterministic guarantees for expected-case performance. 
Less frequently noted in undergraduate treatments is that randomization also functions as an ethical primitive, a tool with value in the unpredictability or indistinguishability it provides.

A modern instance is differential privacy~\cite{dwork2014algorithmic}, where calibrated noise is added to query responses to provide a quantitative privacy guarantee.
The privacy guarantee \emph{is} the randomization: a deterministic mechanism cannot provide it, regardless of design cleverness.
Similar logic applies to randomized fair allocation, randomized auditing, and the use of fair coin extraction~\cite{vonneumann1963various}. 
For each, the randomization is not an inefficiency concession but a substantive component of what makes the system ethically defensible. 
ToC students who have, for example, proved that \textbf{BPP} algorithms can be amplified to arbitrarily small error are positioned to understand the $\varepsilon$, $\delta$ parameters of differential privacy as a quantified bound on a probabilistic guarantee. 
We recommend that when introducing randomized algorithms, present at least one example whose ethical motivation is foregrounded so that students encounter randomization as both a computational and normative tool.

\section{Lightweight Integration Strategies}
\label{sec:strategies}

The most studied integration pattern in the embedded-ethics literature is the short, in-class discussion immediately following a technical result~\cite{grosz2019embedded,skirpan2018ethics}. 
In a ToC course, this maps to a ten-to-fifteen-minute segment after the proof of a major theorem where the instructor poses one or two questions connecting the theorem to a contemporary system or debate; we call this a ``post theorem pause.''
Rice's theorem invites discussion of the verification claims made by deployed program-analysis systems, and the pumping lemma invites discussion of what content-moderation filters can and cannot recognize, etc.
The pattern here is established, requires no philosopher co-instructor, and exploits the fact that students who have internalized a theorem are the audience best positioned to reason from it.

A second pattern, based on value-sensitive design pedagogy and on the integrated activities by \citet{skirpan2018ethics} asks students to write short reflective passages alongside technical exercises. 
In a ToC course, this pattern maps onto the reduction exercises that already constitute problem sets. 
Following any reduction the student constructs, a short commentary asks the student to identify a real-world deployment that depends on the reduction holding, and to name one feature of the situation the reduction does not preserve.
The exercise reuses the reduction the student has already constructed and operates directly on the previous rhetorical-reduction analysis. 
\citet{burton2018teach} report that reflective writing tied to course content engages students across CS contexts. 
We see no reason theory should be an exception.

A third pattern uses a single contemporary news item or deployed system as a recurring point across multiple class sessions so that the technical content accumulates ethical context across the term~\cite{saltz2019integrating}. 
A theory course adopting this pattern might select, at the start of the term, a system whose operation can be illuminated by results from multiple units of the course.
For example, a content-moderation system that admits both formal-language analysis and computability-theoretic critique, or a privacy-preserving analytics system whose guarantees draw on both hardness and randomization. 
The recurring anchor reduces the per-unit cost of integration, since the ethical context need not be re-established at each occurrence, and lets students see the unity of the theoretical apparatus reflected in a single application domain.

The fourth pattern, the most consequential and most often omitted in informal accounts of integration, is the inclusion of ethics content in the instruments by which the course is graded. 
Fiesler et al.'s syllabus analysis~\cite{fiesler2020we} documents wide variation in whether ethics is assessed at all.
Padiyath's realist review~\cite{padiyath2024realist} clarifies the mechanism: when students do not perceive value in an ethics intervention and are graded solely on their performance within it, they tend to satisfy the rubric rather than engage with the underlying questions. 
Assessment is therefore necessary but not sufficient, and the design of the graded component matters as much as its presence. 
We propose that integration in theory courses include at least one assessment touch point per term that asks students to apply a result from the course to an ethical question of the kind raised in the integration segments, rewarding the quality of reasoning rather than adherence to a standard answer key.

\section{Anticipated Objections}
\label{sec:replies}

The time-constraint objection from Section~\ref{sec:obj-time} recurs at the implementation stage, and the strategies of Section~\ref{sec:strategies} are designed against it.
The post theorem pause occupies a class segment and the reduction commentary reuses an exercise.
We note here only that an instructor unwilling to allocate ten minutes per major result to ethical framing has implicitly committed to a view of theory's curricular function that the our position contests. 
Three further objections will be familiar to any instructor who has discussed ethics integration informally with theory colleagues. 
They concern instructor preparation, student reception, and risk of politicization.

\subsection{``This Is Not What I Was Trained To Teach''}
\label{sec:rep-training}

This objection is that theory instructors are mathematicians and theoretical computer scientists, not philosophers, and the suggestion that they lead ethical discussions in their classrooms asks them to perform work for which their training did not prepare them. 

We offer two answers. 
First, the integration patterns of Section~\ref{sec:strategies} do not require the instructor to be the ethical authority in the room. 
The post theorem pause is a question, not a lecture; the reduction commentary is the student's writing, not the instructor's.
The instructor contributes the theorem and the framing question, and the substantive ethical reasoning is done by students, drawing on whatever prior preparation they bring~\cite{reich2020teaching}. 
Second, the worry that one is unqualified to lead ethical discussion is itself a useful modeling moment. 
Students who see their instructor pose a question without claiming to know the answer learn something important about how ethical reasoning differs from theorem-proving, and that difference is part of the content we propose theory courses convey.

\subsection{``Students Will Disengage''}
\label{sec:rep-engage}

This objection holds that students enrolled in a theory course are, by self-selection, oriented toward technical rather than discursive material, and will treat ethical integration as off-task. 
The empirical record does not support this. 
Studies of ethics integration in technically dense CS courses~\cite{saltz2019integrating,skirpan2018ethics} report that integrated modules are received at least as well as standalone ethics offerings precisely because integration makes the connection to technical content visible. 
Padiyath \cite{padiyath2024realist} supplies direct evidence on student attitudes rather than syllabus-level inference: interventions that connect to and challenge students' existing technical commitments generate acceptance, whereas interventions perceived as disconnected from the course's core material are the ones students resist. 
The student who disengages from a context-free ethics module because it is disconnected from other coursework is, on this evidence, the one most likely to engage with an ethics segment that starts with ``recall the theorem we proved last time.'' 
What disengages students is not ethical content but disconnection, and the integrations from Section~\ref{sec:strategies} are designed specifically against that disconnection.

\subsection{``This Risks Politicizing the Course''}
\label{sec:rep-political}

This objection is the most serious.
To raise contemporary ethical questions in the classroom is to invite political controversy into a course whose technical content was likely interpreted to be politically neutral.
The instructor risks alienating students whose political views differ from their own, undermining the authority of the technical material, and converting a ToC course into an arena of partisan dispute.

We accept the underlying concern and reject the conclusion. 
The concern is legitimate: ethics integration done badly (e.g., as advocacy, as one-sided framing, as expectation that students adopt the instructor's conclusions) is a real failure mode documented in the integration literature~\cite{fiesler2020we}, and is corrosive to classroom trust. 
The conclusion, however, conflates \emph{political} with \emph{politicized}. 
To ask what Rice's theorem implies for the verification claims made by deployed systems is political in the broad sense that it concerns the organization of collective life; it is not politicized in the narrow sense of taking sides in a partisan dispute. 
The integration patterns in Section~\ref{sec:strategies} are deliberately structured to keep the distinction operative: they pose questions rather than answers, treat contemporary systems as cases for analysis rather than targets for denunciation, and assess students on the quality of their reasoning rather than the conclusions it reaches. 
An instructor who maintains that distinction is doing work the standalone ethics course cannot straightforwardly do.
They demonstrate in real time what it looks like to apply technical rigor to a politically live question without sacrificing either.

A further dimension of this objection concerns student agency rather than instructor neutrality. 
Padiyath~\cite{padiyath2024realist} reports that students resist ethics interventions framed around scenarios in which they perceive themselves to have little power to act; part of what can make an ethics discussion feel ``political'' is being asked to care about a problem one feels powerless to affect. 
The integration avenues from Section~\ref{sec:avenues} mitigate this by tying ethical questions to decisions the students themselves will make as designers and analysts: which language family a filter uses, what a reduction preserves, where calibrated noise enters a mechanism. 
Where possible, instructors can further ground these segments in concrete decisions that practitioners have actually faced when using these results to build or analyze systems, so that the ethical stakes attach to choices within the students' eventual control rather than to a distant controversy.

\section{Conclusion}
\label{sec:conclusion}

What remains is for our field to act. 
We close with three calls and an empirical research agenda that follows from our position.

\subsection{Calls to Action}
\label{sec:calls}

\paragraph{To theory of computation instructors.} The most consequential move available is the inclusion of a single integration segment of the kind described in Section~\ref{sec:strategies}. 
The ``post theorem pause'' is the lowest-cost option; the reduction commentary is the most distinctive to theory and the one we believe most clearly exceeds what other CS courses can offer. 

\paragraph{To curriculum designers and accreditation bodies.} The CS2023 report~\cite{cs2023} treats Society, Ethics, and the Profession as a knowledge area to be distributed across the curriculum but offers no specific guidance for ToC courses. 
We call on the next revision of the report to include explicit integration recommendations for ToC drawn from the avenues of Section~\ref{sec:avenues} or from following empirical work of this paper. 
The omission of theory from current guidance both reflects and reinforces the disciplinary skew documented in Section~\ref{sec:landscape}, and we recommend specific guidance.

\paragraph{To the computing education research community.} The position of this paper is defended by argument rather than by data. 
We do not claim to have established that the integration patterns we describe will, in practice, produce the outcomes we hope for. 
That is empirical work, and needs to be done.

\subsection{An Empirical Research Agenda}
\label{sec:agenda}

We identify four research questions that follow directly from the
position and that the field is now positioned to address.

\begin{enumerate}
    \item What is the current state of ethics integration in theory courses (e.g., a targeted syllabus analysis similar to \citet{fiesler2020we})?

    \item Which of the five avenues of Section~\ref{sec:avenues} produces the strongest student engagement and learning outcomes? A comparative study is warranted.  

    \item Do theory students who experience integration segments demonstrate transfer to ethical reasoning in subsequent technical courses? 
    The deepest claim of the embedded-ethics movement is that integration produces graduates who reason ethically about the systems they build; this to our knowledge has not been tested for theory.

    \item Do the objections of Section~\ref{sec:replies} materialize in practice?
    The objections are reported informally by instructors and rebutted here by argument. 
    Whether students disengage, courses become politicized, and instructors find the work tractable are important empirical questions.
\end{enumerate}

The position of this paper is that ToC belongs in the embedded-ethics movement. 
The empirical work above is what would move that position from defensible to confirmed. 
We invite the community to undertake it.

\begin{acks}
We thank the anonymous reviewers for their helpful comments.
The opinions in this work are solely of the author, and do not necessarily reflect those of the U.S. Army, U.S. Army Research Labs, the U.S. Military Academy, or the Department of Defense.
\end{acks}

\bibliographystyle{ACM-Reference-Format}
\balance
\bibliography{sample-base}

\end{document}